\documentclass[12pt,preprint]{aastex}
\slugcomment{(revised February 2004)}
\shorttitle{HIGH-VELOCITY BROAD LINE REGION} \shortauthors{SNEDDEN
AND GASKELL}
\begin{document}
\title{The Case for Optically-Thick High Velocity Broad Line Region Gas
in Active Galactic Nuclei}
\author{Stephanie A. Snedden\altaffilmark{1} \and C. Martin Gaskell}
\affil{Department of Physics and Astronomy, University of
Nebraska, Lincoln, NE 68588-0111} \email{snedden@apo.nmsu.edu}
\email{mgaskell1@unl.edu}
\altaffiltext{1}{Current address: Apache Point Observatory, 2001
Apache Point Rd., Sunspot, NM 88349-0059}
\begin{abstract}
A combined analysis of the profiles of the main broad quasar
emission lines in both {\it Hubble Space Telescope} and optical
spectra shows that while the profiles of the strong UV lines are
quite similar, there is frequently a strong increase in the
Ly$\alpha$/H$\alpha$ ratio in the high-velocity gas. We show that
the suggestion that the high velocity gas is optically-thin
presents many problems. We show that the relative strengths of the
high velocity wings arise naturally in an optically-thick BLR
component. An optically-thick model successfully explains the
equivalent widths of the lines, the Ly$\alpha$/H$\alpha$ ratios
and flatter Balmer decrements in the line wings, the strengths of
C\,{\sc III}] and the $\lambda$1400 blend, and the strong
variability of high-velocity, high-ionization lines (especially
He\,{\sc II} and He\,{\sc I}).
\end{abstract}
\section{Introduction}
The nature and origin of the broad-line region (BLR) in active
galactic nuclei (which we will refer to simply as ``quasars'') has
been a long-standing problem. It has been known for some time that
the shapes of broad emission lines differ from line to line within
the same object depending on the ionization level of the species
producing the each line.  For example, He\,I $\lambda$5876 is
broader than H$\alpha$ (Osterbrock \& Shuder, 1982), O\,I
$\lambda$1304 and C\,II] $\lambda$2326 are narrower than C\,IV
(Wilkes 1984), C\,IV $\lambda$1549 tends to be broader than Mg\,II
$\lambda$2798 (Mathews \& Wampler, 1985) and H$\alpha$ is narrower
than H$\beta$ (Osterbrock 1977). H$\alpha$ and H$\beta$ have
weaker wings than Ly$\alpha$ (Zheng 1992; Netzer et al.
1995)\footnote{It can be misleading to use FWHM as a measure of
how broad emission lines are in the wings. Large contributions to
the core can skew the FWHM to lower values. For example, although
the line profile ratios reveal that the Ly$\alpha$ wings are
broader, Table 1 of Netzer et al. (1995) shows that FWHM$_{H
\beta}
> FWHM_{Ly\alpha}$}. These differences in line profiles imply differences
in physical conditions as a function of velocity and probably as a
function of distance from the central ionizing source. One of many
unresolved BLR questions is whether these differences can be
explained by one BLR with a range of conditions as a function of
radius, or whether there are two (or more) fundamentally distinct
components. It is common to discuss two separate components, but
there are differences in the literature as to how to divide the
BLR into two components.  In disk-plus-wind models (e.g., Chiang
\& Murray 1996; Bottorff et al. 1997) the two components are
identified with emission from near the disk and emission from a
wind above the disk. Some observers and phenomenologists have made
distinctions primarily on the basis of the mean degree of
ionization, others primarily on the basis of Doppler widths. There
is substantial overlap in these classifications. Gaskell (1987)
and Collin-Souffrin \& Lasota (1988) make a division, motivated by
profile differences and photoionization modelling issues, into two
components, one with a typical nebular spectrum (called ``BLR I''
by Gaskell 1987, and ``HIL'' by Collin-Souffrin \& Lasota) and one
with substantial emission from clouds with large partially-ionized
zones (``BLR II'' or the ``LIL'').  The case for two such
components is summarized by Gaskell (2000).  Other workers,
motivated by analyses of line profiles, have separated the BLR
into a ``very broad line region'' (VBLR) and another component.
The other component has been called the ``intermediate-line
region'' (ILR) by Wills et al. (1993) and Brotherton et al. (1994)
or the ``classical broad component'' (BC) by Sulentic, Marziani,
\& Dultzin-Hacyan (2000).  Note that the BC and ILR are not
identical and hence the corresponding VBLR components can be quite
different. The VBLR, ILR, and BC show both high and low ionization
emission. Despite these differing terminologies, there is general
agreement that the conditions of the highest velocity gas (which
we will call the VBLR) are different from the other components
although the VBLR gas might merely be an extension of the other
gas. Historically, photoionization models assumed that
line-emitting gas was optically-thick. However, several authors
have argued for a significant optically-thin component to the VBLR
gas, and to date, there has been no resolution to this debate. The
main arguments for an optically-thin VBLR have been:
\begin{enumerate}
\item {\it The general symmetry of Ly$\alpha$.} Wilkes \& Carswell
(1982) pointed out that BLR clouds could not be both
optically-thin and have a net radial motion (e.g., outflow)
because Ly$\alpha$ profiles are symmetric and similar to C\,IV
profiles.  Since the blueshifting of the high-ionization lines
requires at least some radial motions (Gaskell 1982), the relative
symmetry of Ly$\alpha$ could mean that the VBLR (highest velocity
gas associated with the profile wings) has to be optically-thin.
\item {\it An apparent lack of variability of the VBLR in Mrk~590
between two epochs.} Ferland, Korista, \& Peterson (1990) found
that the VBLR Balmer emission line flux in Mrk~590 changed little
between two epochs about three years apart even though the
continuum and line cores changed. Hydrogen recombination lines
should change little if the VBLR is optically-thin and hence fully
ionized. \item {\it The Ly$\alpha$/H$\beta$ ratio is higher for
the VBLR}. Zheng (1992) showed that for some AGNs the
Ly$\alpha$/H$\beta$ ratio is much higher in the wings and that it
approached the Menzel-Baker case B value that might be expected
from optically-thin gas. \item {\it Emission line responses might
require negative responsivities.} Sparke (1993) pointed out that
the shapes of some of the line-continuum cross-correlation
functions from 1989 monitoring of NGC\,5548 seem to require that
the emission of the inner BLR declines as the continuum increases.
She suggested that optically-thin clouds in the inner BLR could do
this.
\end{enumerate}
The advent of the {\it Hubble Space Telescope} (HST) now makes
possible the comparison of high-quality line profiles of both the
high-ionization UV lines and the Balmer lines in the same objects.
In this paper we use such comparisons to investigate the nature of
the VBLR and we will argue that the bulk of the VBLR is in fact
{\it not} optically-thin.  In paper II (Snedden \& Gaskell, in
preparation) we discuss the physical conditions in the BLR gas as
a function of velocity.
\section{DATA ANALYSIS}
We have determined the Ly$\alpha$/H$\alpha$ ratio as a function of
velocity for eight AGNs with high signal-to-noise HST FOS and
optical spectra. Stirpe (1990, 1991) presented high-quality Balmer
line profiles for bright low-redshift AGNs.  For this study we
have chosen eight of these that also had high signal-to-noise HST
FOS spectra available. Six out of these eight AGNs are radio
quiet.
\subsection{Flux Levels}
The objects in the sample were not observed simultaneously in the
UV and optical bandpasses, however this  only introduces an
uncertainty in the ratios at the 20 percent level and does not
affect our conclusions. As a check of the effect of variability on
our sample, the line ratios of seven of our objects were compared
with integrated line fluxes from the literature that were measured
at different times. Six of these were found to be the same to
within 20 percent. Another, PKS 2251+113, showed a 20 percent
change in flux compared with Espey et al. (1994) and a 30 percent
change compared with Boroson and Green (1992). The remaining
object (B2 2201+315=4C\,31.36) had been subject to long-term
broad-band optical monitoring and optical continuum variations
were less than 5 percent. Line variations will be less than this.
We also compared {\it UV} lines of NGC 5548 at high and low states
during the International AGN Watch campaign with lines in the
optical spectra taken at one (different) epoch, and found only a
10 percent difference in the predicted physical conditions of the
BLR gas. Since the optical spectra were obtained with a narrow
slit, we determined and applied a small (17 percent) correction to
the optical line fluxes. This value was obtained by comparing the
Stirpe sample's integrated Balmer line fluxes with those in the
literature and has a negligible effect on our results here.
\subsection{Line Profiles}
    Determining the profiles of the broad emission lines in AGNs is difficult.  Two issues
    that particularly affect the line profile are (i) subtraction of the appropriate continuum
    level and (ii) de-blending of weaker lines.

\subsubsection{ The Effects of Continuum Level Placement}

    Errors in line profiles due to continuum-level placement only have a significant effect
    in the high-velocity wings of the lines.  In Figure 1\emph{(a)}, \emph{(b)} we show a comparison of
    Ly$\alpha$
    and H$\alpha$ for a typical object.  In this paper we will only be deriving physical conditions at
    five projected velocities: 0 km s$^{-1}$, $\pm 5000$ km s$^{-1}$, and $\pm
    10,000$ km s$^{-1}$.  It can be seen that
    for Ly$\alpha$, even at $\pm 10,000$ km s$^{-1}$, the effect of errors in the continuum placement are
    only $\sim 20$ percent in most objects, which is comparable to the uncertainty due to counting statistics
    and unimportant for our analyses.
    The wings of the Balmer lines are considerably weaker in all our objects and, as can be
    seen from Figure 1, the continuum placement is clearly more critical.  The most problematic
    line is H$\beta$ and Shuder (1982) in his Figure 2 has illustrated the effects of different
    choices of local continuum.  The line for which it is easiest to
define an underlying continuum is H$\alpha$, but, as can be seen
in our Figure 1\emph{(a)}, the H alpha flux at $\pm 10,000$ km
s$^{-1}$ can be uncertain by a factor of two.  In general
H$\alpha$ is very symmetric in our objects and the consistency we
get in flux ratios between the two wings for our sample as a whole
suggests that possible errors in the choice of the continuum are
not a major issue.

\subsubsection{De-blending}

    The degree of blending varies from line to line.  The least problematic line is H$\alpha$.
    It is only blended with the narrow [N II] lines which are easy to deal with.  Apart from
    occasional narrow absorption lines, Ly$\alpha$ is free of blending with other emission
    lines on the short wavelength side but on the long wavelength side it is blended with
    N V $\lambda$1240 and Si II $\lambda$1265.  These blends cover our +5000 to +10000 km s$^{-1}$ velocity range.
    The line with most serious blending problems is H$\beta$.  The narrow but strong [O III]
    lines occur at +6000 and +9000 km s$^{-1}$ in the red wing of H$\beta$, and the most problematic
    blend is the broad He II $\lambda4686$ which occurs at -11,000 km s$^{-1}$ and therefore contributes
    significantly to our -10,000 km s$^{-1}$ region.  Many weak Fe II lines can also be a factor.
    The effects of uncertainties in continuum placement and de-blending for H$\beta$ have been
    extensively investigated by Shuder (1982, 1984) and Crenshaw (1986).

    The de-blending of the
    Balmer lines in the objects we consider was performed by Giovanna Stirpe and is
    discussed in detail
    in Stirpe (1990) where raw and de-blended profiles can be seen for all objects.
    We de-blended the Ly$\alpha$ region with multiple Gaussians using the SPECFIT and
    SPLOT tasks in IRAF.  The rest wavelengths were fixed but the line intensities and widths
    were unconstrained.  We used one or two Gaussians per line.  This method necessarily assumes
    that the lines are symmetric.  A real excess in the red wing of Ly alpha is hard to
    distinguish from N V and Si II emission.  This is a fundamental limitation of {\it all}
    de-blending.  In Figure 1\emph{(a)},\emph{(b)} we show the Ly$\alpha$ region {\it with no de-blending at all}.
    This represents a worst case scenario for de-blending.  It can be seen that while the overall
    profiles of H$\alpha$ and Ly$\alpha$ are fundamentally different, there is good agreement in the
    short wavelength wings where de-blending is not an issue.  On the red side, where blending with
    N V and Si II is a problem, the profiles disagree somewhat.  Although the true
    Ly$\alpha$
    profile on the red side is unknowable, it cannot be higher than the observed profile.  Since we
    know that there {\it is} N V and Si II in AGNs, e.g. Vanden Berk et al. (2001), the true Ly$\alpha$ profile in the red wings
    must be lower than the observed.  As can be seen from Figure 1, the {\it maximum} error in the red
    wing is therefore about a factor of two.  The consistency of the physical conditions we deduce for the
    red and blue wings suggests that the error in de-blending the red wing of Ly$\alpha$ is much
    less than a factor of two.

\subsection{Reddening}
Since we consider Ly$\alpha$/H$\alpha$ ratios, reddening effects
are significant, and corrections are required. For Galactic
reddenings we adopted the average of the reddenings given by
Burstein and Heiles (1982) and Schlegel et al. (1998). We give
these in Table 1. In addition, there is probably internal
reddening in the AGNs themselves. De Zotti \& Gaskell (1984) found
typical reddenings of E(B-V) $\sim 0.25$ for Seyfert 1 galaxies.
Gaskell et al. (2004) have shown that the {\it nuclear} reddening
curve for AGNs is flat in the UV.  They find typical reddenings of
E(B-V) from 0.1 -- 0.4 in the optical. These ranges agree with
E(B-V) $\sim 0.25$, as found by Ward \& Morris (1984) for several
line ratios of NGC\,3783 (i.e., 0.13 greater than the Galactic
reddening in Table 1). Netzer et al. (1995) and Bechtold et al.
(1997) found that the Ly$\alpha$/H$\alpha$ ratio for quasars was
correlated with the 0.1 -- 0.6 $\mu$m flux ratio. For a typical
quasar this implies a reddening of E(B-V) $\sim 0.15$ (including
the Galactic contribution).  We therefore believe that the total
reddenings could have a Galactic reddening equivalent of 0.1 or
more greater than those given in Table 1 (assuming a standard
Galactic reddening curve). In Table 2 we show Galactic
reddening-corrected Ly$\alpha$/H$\alpha$ ratios at $\pm 10000$
km~s$^{-1}$, $\pm 5000$ km~s$^{-1}$, $\pm2500$ km~s$^{-1}$ and 0
km~s$^{-1}$ for our sample\footnote{Note that NGC\,3783 for which
Ward \& Morris (1984) get a higher reddening, has the lowest
Ly$\alpha$/H$\alpha$ ratio.}. The ratios are averaged over $\pm
1000$ km~s$^{-1}$. In Table 3 we show the H$\beta$/H$\alpha$
ratios at the same velocities. The most uncertain values are those
at +10,000 km~s$^{-1}$ where N\,{\sc V} causes problems for
Ly$\alpha$ and [O\,{\sc III}] causes problems for H$\beta$. Plots
of the Ly$\alpha$/H$\alpha$ ratio profiles for three individual
objects can be seen in Snedden \& Gaskell (1997, 1999b) but these
plots were not reddening corrected and the vertical scale was
calculated incorrectly (it is too large by a factor of 6563/1216 =
5.4).  In Figs. 2$a$ and 2$b$ we show the average
Ly$\alpha$/H$\alpha$ and H$\alpha$/H$\beta$ ratios for our entire
sample as a function of velocity.  These values are taken from
Tables 2 and 3. To illustrate the uncertainty due to the reddening
correction we show the effect of dereddening by an additional
E(B-V) = 0.1 as smooth curves. It is possible in some objects the
actual dereddening needed could be even greater, although Gaskell
et al. (2004) find that the average reddening of high-luminosity
AGNs is low.

\section{PHOTOIONIZATION MODELS}
It has long been known that the hydrogen density, $n_H$, and the
ionization parameter\footnote {The ionization parameter is defined
as $U = \Phi~(H) \div (n(H) \cdot c$), where $\Phi(H)$ is the
surface flux of ionizing photons ($cm^{-2} s^{-1}$), $n(H)$ is the
total hydrogen density (ionized, neutral and molecular), and $c$
is the speed of light.}, $U$, of quasar broad line regions can be
determined from the relative strengths of the C\,IV $\lambda
1549$, Ly$\alpha$, and C\,III] $\lambda 1909$ lines (e.g.,
Davidson 1977, Mushotzky \& Ferland 1984).  We have generated
grids of line intensity ratios for these emission lines from
models calculated using the photoionization code CLOUDY, version
C90.04 (Ferland 2002). In CLOUDY ionization equilibrium and
thermal equilibrium are solved for in zones of approximately
constant conditions.  On the side of the cloud where the ionizing
radiation enters, hydrogen is essentially completely ionized. When
ionizing photons are no longer reaching a zone, the fraction of
neutral hydrogen quickly rises (see Osterbrock 1989). This is the
classical Str\"{o}mgren length.  The presence of X-rays in an AGN
makes the transition from hydrogen being fully ionized to being
largely neutral less abrupt in a BLR cloud than in a classic H II
region or planetary nebula.  We have therefore defined the
Str\"{o}mgren length in a BLR cloud to be where the ionization
fraction of hydrogen falls to 95 percent (Snedden \& Gaskell
1999a). In CLOUDY C90 this corresponds to about 200 zones.  In the
models present here we used the standard quasar continuum of
Ferland (2002). For all calculations solar abundances were used.
As we have shown in Snedden \& Gaskell (1999a), enhanced
abundances change the ionization structure, with the result that a
given C\,III]/C\,IV ratio corresponds to a higher density.  This
does not change any of the conclusions presented here.  We ran
grids of both optically-thick and optically-thin models. Other
details of the models are as in Ferland \& Persson (1989).

The optically-thick calculations were stopped at a column density
of $10^{25.5}$ as in Ferland \& Persson (1989). To see how line
ratios behaved in the optically-thin limit we ran a model that
stopped after only three zones (Lyman continuum optical depth,
$\tau_{912} \sim 10^{-4}$). Although illustrative, such an extreme
model has so little gas that even with a 100 \% covering factor it
cannot produce significant emission.  In order to produce the
minimum observed $W_{\lambda} = 50$\AA\, with a covering factor of
50 \% we find that we need $\tau_{912} \sim 1.0$.  We therefore
ran a third set of models that stopped at $\tau_{912} = 1$. In
Figures 3 and 4 we show grids of the ratios of
Ly$\alpha$/C\,IV/C\,III] and Ly$\alpha$/H$\alpha$/H$\beta$ as
functions of $U$ and $n_H$ for optically-thin, optically-thick and
$\tau_{912} = 1$ clouds. It should be noted that there can be
uncertainties in line and continuum transfer that are difficult to
estimate, since CLOUDY uses escape probabilities instead of more
formally correct calculations (Ferland 2002). At the high
densities and pressures of the BLR, this can be a serious source
of error. Fortunately, the approximation is known to be exact if
physical conditions are held constant over the region of interest.
Since we consider emission line ratios as a function of velocity,
rather than ratios integrated across the velocity field, we meet
this criterion. As has long been recognized (e.g., Davidson 1977),
most of the strong UV lines in quasar spectra can be explained by
a ``canonical'' density of $10^{10}$ cm$^{-3}$ and canonical
ionization parameter of the order 0.1. ``Locally
Optimally-emitting Cloud'' (LOC) scenarios (Baldwin et al., 1995;
see also Bottorff \& Ferland 2001) can fit many line ratios by
combining different clouds with different properties, but by
considering a single zone and the major recombination lines, we
can compare optically-thick and optically-thin clouds
unambiguously. Furthermore, the simplest case, that of single-zone
models, matches these strong lines well.

\section{HYDROGEN LINE RATIOS}
Zheng (1992) suggests that an optically-thin component is one way
of getting a higher Ly$\alpha$/H$\alpha$ ratio in the wings. As
discussed in \S 3, our very optically-thin model fails to
reproduce the observed $W_\lambda$. Furthermore, from Figure
3\emph{a} and the Ly$\alpha$/H$\alpha$ ratios in Table 2, it can
be seen that the ratios predicted by this thin-limit model are in
fact too high for both the wings and core, and could only be
explained by either extremely high values of $U$, or extreme
intrinsic reddening. If we use the flat nuclear reddening curve
found by Gaskell et al. (2004), E(B-V) $\sim 1.0$ would be
required. The Balmer decrements in the three-zone optically-thin
model are very uniform (close to case B) and do not match the
observed values in the wings where the decrement is quite flat
(i.e., H$\beta$ is too strong), unless the errors in continuum
subtraction are at the 100 percent level, the high end of Stirpe's
(1991) error estimate. As has been noted by Shuder (1982, 1984)
and Crenshaw (1986), determination of the flux in the wings of
H$\beta$ is very uncertain. If the continuum is indeed set too
low, we can match the thin limit model predictions in the wings
but cannot match the Ly$\alpha$/H$\alpha$ ratio assuming the same
physical conditions. If the error in the Balmer decrement is in
fact less than a factor of 2, we cannot match any of the observed
hydrogen lines to the three-zone model.

 The $\tau_{912} = 1$ models fit the
Ly$\alpha$/H$\alpha$ ratios in the high velocity wings and also
match the lowest observed $W_\lambda$ if we assume a large
covering factor (50 percent). However, they cannot match the
observed ratios for the line cores unless reddening is extreme.
These models are, of course, close to being optically thick.

The optically-thick model is the best match to both observed line
ratios and $W_\lambda$ over the entire emission line profile from
-5000 km s$^{-1}$ to +5000 km s$^{-1}$. The oval in Figure
3\emph{b} shows the region in physical parameter space that
matches the observed emission line ratios. We note that it is not
possible to simultaneously match Ly$\alpha$, H$\alpha$ and
H$\beta$ at $\pm 10000$ km s$^{-1}$ with \emph{any} set of models.
We believe this could be a result of an over-subtraction of the
continuum at H$\beta$.

We cannot rule out a contribution to the wings from gas with
$\tau_{912} \approx 1$, but we reiterate that this set of models
requires a large covering factor and in fact approaches the
optically-thick case. Furthermore, we will show in \S 7 that these
same models lead us to a similar conclusion regarding the C III]
and C IV emission lines. In this case, the photoionization models
support an optically-thick ILR and a  VBLR that could have an
optically-thin component, or be optically-thick. When the overall
symmetry of the ILR component of Ly$\alpha$ is considered (see
Figure 1 of Brotherton et al. 1994), it is reasonable to conclude
that there is a negligible net radial motion of the
optically-thick gas. This is consistent with the results of
reverberation-mapping studies (Gaskell 1988, Koratkar \& Gaskell
1991), and a necessary assumption in using such studies to
calculate black hole masses.

We believe the simplest explanation is that the gas responsible
for the high-velocity wings of the hydrogen emission lines
contains a significant optically-thick component, although some
optically-thin gas could also be present.

\section{PROFILES OF LOW-IONIZATION LINES}
H$\beta$ is produced both as a recombination line in the
highly-ionized zone of clouds and as a collisionally-excited line
in the partially-ionized zone.  An optically-thin cloud lacks the
partially-ionized zone and will not produce low-ionization
emission lines such as Fe\,II, Mg\,II, O\,I, etc. If we assume
that a high-ionization line includes an optically-thin,
high-velocity component, the profile of another line arising only
from the optically-thick, partially-ionized zone will have a
different, narrower profile. Unfortunately there is no strong
uncontaminated low-ionization line for comparison with H$\beta$.
The strongest low-ionization line is Mg\,II~$\lambda$2798.
Comparisons of the wings of Mg\,II and H$\beta$ are complicated by
the Fe\,II emission around Mg\,II but Gaskell \& Mariupolskaya
(2002) find the FWHMs of Mg\,II and H$\beta$ to be well-correlated
for all radio classes. They find no systematic difference in the
FWHMs. Detailed profile comparisons (Grandi \& Phillips 1979) are
consistent with Mg\,II and H$\beta$ having identical profiles.
Although much weaker than Mg\,II, and in a spectral region much
affected by telluric features, O\,I~$\lambda$8446 is relatively
free from contamination.  Morris \& Ward (1989) made profile
comparisons between H$\alpha$ and O\,I~$\lambda$8446 to search for
the presence of optically-thin material.  Although their
comparisons have been quoted as supporting the existence of
optically-thin material, the majority of their objects have
identical H$\alpha$ and O\,I~$\lambda$8446 profiles and they state
that ``optically-thin material contributes a negligible amount of
the H$\alpha$ material''. Their two clearest cases of profile
differences are both objects showing complex line structure that
we have argued elsewhere (Gaskell \& Snedden 1999) is caused by
emission from a disk component.  The small differences between the
H$\alpha$ and O\,I~$\lambda$8446 profiles clearly arise from this
disk component. It must have conditions producing a slightly
different H$\alpha$/O\,I~$\lambda$8446 ratio. Rodriguez-Ardila,
Pastoriza, \& Donzelli (2000) find similar H$\alpha$ and
O\,I~$\lambda$8446 profiles for all but one of the AGNs they study
and they also conclude that optically-thin gas with a different
velocity field (e.g., a VBLR) contributes a negligible amount of
H$\alpha$ emission. We conclude that there is no convincing
evidence from the profiles of low-ionization emission lines for a
significant contribution to the line flux from optically-thin gas.

\section{VARIABILITY OF HELIUM LINES}
Although lack of variability of the wings of the Balmer lines
between two epochs was the motive for Ferland et al. (1990)
proposing that the highest velocity gas was optically-thin,
difference and root-mean-square (rms) spectra show that the most
variable broad line component is the {\it high}-ionization
component. This can be seen, for example, in the rms spectrum of
NGC\,7469 (see Figure 2 in Collier et al. 1998) and in many other
rms spectra (e.g., see Figures 1 -- 3 in Peterson et al. 1998).
The variable component of He\,II~$\lambda$4686 is always very
broad, making it a quintessential VBLR line.  In the UV
He\,II~$\lambda$1640 is similarly variable (see, for example,
Krolik et al. 1991). Peterson et al. (1990) show that most of the
variability of both the high-ionization UV lines and the Balmer
lines in NGC\,5548 is due to the broad, slightly blueshifted
component\footnote{This variability should not be confused with
object-to-object profile differences that tend to be in the {\it
lower}-velocity gas (see Francis et al. 1992 and Brotherton et al.
1994).}. In an optically-thick cloud most of the variation in the
line flux of a recombination line is caused by changes in the size
of the ionized regions as the ionizing flux varies. An
optically-thin cloud produces only small changes in the line flux
of a recombination line from a fully-ionized ion (such as H\,I and
He\,II) because the Str\"{o}mgren length of the ionized regions is
larger than the cloud size.  Such lines actually {\it decline}
slightly in flux as $U$ increases because the recombination
coefficients decline at higher temperatures. The theoretical lack
of variability of He\,II in an optically-thin cloud is clearly
shown in Figure 8$b$ in Shields, Ferland, \& Peterson
(1995)\footnote{The slight rise of He\,II at low $U$ in their
Figure 8$b$ is because their ``thin'' model is thicker than ours
and their He$^{++}$ zone does not extend to the back of the cloud
for low $U$.}. Optically-thick models, on the other hand, produce
strong He\,II variability, as is seen. Clouds that are
optically-thin in the Lyman continuum but with a He$^{++}$
Str\"{o}mgren length less than the cloud size will still produce
strong He\,II variability, but the He$^{++}$ comes at the expense
of the He$^{+}$ so He\,I~$\lambda$5876 would decline as the
continuum brightens.  This has never been reported. Instead, like
He\,II~$\lambda$4686, He\,I~$\lambda$5876 line is a highly
variable VBLR line (see Figure 2 in Collier et al. 1998). Because
of the strong variability of the He\,I and He\,II lines we believe
that the VBLR is not predominantly optically-thin.

\section{THE C\,{\sc III]}/C\,{\sc IV} RATIO}
 Brotherton et al. (1994) noted that observed values of
C\,III]/C\,IV in the VBLR were higher than expected from their
photoionization models. Our optically-thin carbon line model grid
(Figure 4$a$) shows that increasing $U$ actually leads to a
\textit{decrease} in C\,III]/C\,IV, compounding the discrepancy
between observed and modelled ratios for the VBLR. However, our
optically-thick grid shows an insensitivity to changes in the line
ratio as $U$ increases. The lowest optical depth model that can
replicate the observed $W_\lambda$ of the broad component of
Ly$\alpha$ (Figure 4$c$) approaches being optically-thick with
respect to the carbon
 line forming gas. As mentioned in \S 4, this $\tau_{912} =1 $
 model allows for a combination of optically-thin and thick clouds
 to generate the observed Ly$\alpha$/H$\alpha$ ratio, however, only
 models of optically-thick gas can reproduce the observed carbon line
 ratios over a wide range of $U$. Without high values of $U$ it is difficult to
 reconcile the observed broad width of the C\,{\sc III]} line with photoionization
 models' predictions, at least for high luminosity AGN. Furthermore, the C\,III]/C\,IV ratios
 Brotherton et al. (1994) report for
their ALS VBLR sample match the values for our optically-thick
model grid, yielding reasonable values of $U=0.05$ and
$n_H=5.0\cdot 10^9$. Even if the C\,{\sc III]}/C\,{\sc IV} ratio
in Brotherton et al. is too high as a result of incomplete
deblending, the optically-thick models still allow for higher
values of $U$ for any reasonable value of C\,{\sc III]}/C\,{\sc
IV}. This argues that the VBLR is better modelled with
optically-thick clouds, although we cannot discount a contribution
from optically-thin gas further from the central engine. We do not
expect this component to contribute much to the high-velocity
wings.

\section{THE $\lambda$1400 BLEND}
Optically-thin and optically-thick models predict very different
(Si\,IV+O\,IV])/C\,IV ratios. In the optically-thin case, over the
canonical ranges of $U$ and $n_H$, the emission from both Si\,IV
and O\,IV] are negligible compared to that of C\,IV, yielding only
(Si\,IV+O\,IV])/C\,IV = 0.02. One would not expect to see any
contribution from the VBLR to the $\lambda$1400 blend in observed
spectra if the VBLR component is optically-thin. Brotherton et al.
(1994), however, give a value of 0.42 for the
(Si\,IV+O\,IV])/C\,IV ratio for the ALS VBLR. Although our solar
abundance optically-thick model does not quite yield this ratio,
its value of 0.28 is much closer than the optically-thin model's.
Although we do not address the effect of enhanced abundances on
all line ratios in this paper, we point out that the ratio
measured by Brotherton et al. can be fit with a 5$Z_\sun$
optically-thick model.
\section{DISCUSSION}
\subsection{Variability of Emission from High-Velocity Gas}
Following Ferland et al. (1990), a purported lack of variability
of the VBLR flux in response to continuum variation has often been
quoted as an argument for an optically-thin VBLR.  However, the
Mrk 590 variability Ferland et al. discuss is between only two
epochs separated by three years. Ferland et al. mention additional
examples by Gondhalekar (1990), O'Brien, Zheng, \& Wilson (1989),
and Perez, Penston, \& Moles (1989).  The first two papers
consider low signal-to-noise ratio IUE spectra.  Gondhalekar
(1990) considers the C\,IV and Ly$\alpha$ variability of a number
of quasars. While the difference spectra in some cases do seem to
show a narrower peak to C\,IV, given the poor signal-to-noise
ratio (which is even worse in the difference spectra) and
uncertainties in continuum fitting, it is not clear how
significant these are. There are also about an equal number of
cases of the difference spectra being broader than the individual
spectra. O'Brien, Zheng, \& Wilson (1989) analyze IUE spectra (and
some H$\beta$) spectra of 3C\,273. Given the low signal-to-noise
ratio it is not clear whether the wings of the line really do vary
less. Perez, Penston, \& Moles (1989) do show some cases of the
core appearing to vary more than the wings, but the line in
question is Mg\,II so this cannot be taken as support for an
optically-thin VBLR since Mg\,II arises from optically-thick
clouds. The more extensive reverberation-mapping campaigns of the
last decade and a half give better insights into line profile
variability. Two relevant results for the issue of VBLR
variability are:
\begin{enumerate}
\item The lag for high-ionization lines is shorter than for
lower-ionization lines because the gas closer to the center of the
AGN is more highly ionized (Gaskell \& Sparke 1986, Krolik et al.
1991, Korista et al. 1995, Kriss et al. 2000, etc.). \item There
are long-term changes in the profiles of lines.  These changes are
unrelated to changes in the ionizing flux (Wanders \& Peterson
1996).
\end{enumerate}
These results mean that ({\it i}) in comparing profiles allowance
must be made for the differing time delays and ({\it ii}) in
looking for profile changes due to changing ionizing continuum
levels caution is needed when comparing spectra separated by much
more than the light crossing time.  We believe that failure to
allow for these two effects vitiates the result of the profile
comparisons quoted in Ferland et al. (1990).  For NGC\,5548, for
example, the He\,II lag is an order of magnitude smaller than the
H$\beta$ lag (Korista et al. 1995). Examples of the significant
improvements in emission-line results when lags are allowed for
can be found in Pogge \& Peterson (1992) and Shields et al.
(1995). Gaskell (1988) analyzed the extent of the high-velocity
C\,IV wings in NGC\,4151 in the well-sampled, short-term IUE
monitoring campaign of Ulrich et al. (1984) and found that the
extent of the wings correlated well with the continuum flux.  It
is clear that in NGC\,4151 the VBLR is the most variable C\,IV
component (see Figure 1 in Ulrich et al. 1984 or Figure 1 in
Stoner et al. 1984). This agrees with the He\,II and He\,I VBLR
variability discussed above. rms-difference spectra offer a
powerful test of whether the wings vary less than the cores of
lines.

Many AGNs have been subject to high-quality monitoring, especially
in the optical, and many rms-difference spectra are available. The
key issue is whether the line wings are broader or narrower in the
rms-difference spectra compared with the mean spectra. For the 17
AGNs considered by Kaspi et al. (2000) the FWHMs of the mean and
the rms-difference spectra are very well correlated. We find an
even tighter correlation for the FWHMs of the 19 objects given by
Wandel, Peterson, \& Malkan (1999). For the latter sample we find
the FWHM$_{rms}$/FWHM$_{mean} = 1.07 \pm 0.4$ with FWHM$_{rms}$
being on average 230 km~s$^{-1}$ wider. For the Kaspi et al.
(2000) sample we find FWHM$_{rms}$/FWHM$_{mean} = 0.89 \pm 0.7$.
Gaskell \& Mariupolskaya (in preparation) made independent
measures of FWHM$_{rms}$ and FWHM$_{mean}$ for half the Wandel et
al. sample and their FWHM$_{rms}$/FWHM$_{mean}$ ratio is
consistent with what we get from the Wandel et al. measurements.
For the two objects in common between Wandel et al. (1999) and
Kaspi et al. (2000) the FWHM$_{mean}$ measurements agree well, but
Kaspi et al. have FWHM$_{rms}$ too small by 1000 km~s$^{-1}$. This
suggests systematic differences in how FWHM$_{rms}$ is measured.
Although FWHM$_{mean}$ and FWHM$_{rms}$ are the same on average,
for individual objects the differences can be greater than the
measuring errors.  We propose that these differences result from a
combination of changes in the line flux due to ionizing continuum
changes and the longer-term changes in line profiles due to other
factors. It is worth noting that both the mean and rms-difference
profiles are different for differing observing seasons (see
Wanders \& Peterson 1996). The FWHM does not, of course,
necessarily reflect differences in the wings. Inspection of the 9
mean and rms-difference H$\beta$ spectra shown in Wanders (1997)
shows that the high-velocity wings in most objects are as variable
as the cores, although He\,II introduces uncertainty on the blue
side and [O\,III] on the red.  Mrk 590 in particular has
high-velocity wings present in the rms-difference spectrum. We
conclude that rms-difference spectra strongly support high
variability of the VBLR.

\subsection{Negative Responsivity?}
Sparke (1993) has proposed that the innermost gas in NGC\,5548 is
optically-thin and has a negative responsivity.  This potentially
explains three curious results of the 1989 monitoring campaign:
viz., that some of the line auto-correlation functions are broader
than the continuum autocorrelation function, that some of the
line-continuum cross-correlation functions rise faster than the
continuum auto-correlation function and that the H$\beta$ transfer
function does not peak at zero lag. Although this is an
interesting idea, we do not believe it is tenable because detailed
photoionization models are unable to reproduce the required
negative responsivity for the lines while producing enough line
flux.  When a cloud becomes optically thin (e.g., when the
ionizing flux increases so that the Str\"{o}mgren length exceeds
the size of the cloud) the L$\alpha$ flux ceases to rise. The He
II line flux will continue to rise for another order of magnitude
in $U$ as the size of the He$^{++}$ zone increases. When the
He$^{++}$ zone equals the cloud size the He II flux will also
cease to increase.  C IV line flux rises with the He II flux, but
while the He II flux plateaus, the C IV flux will fall rapidly
after a further order of magnitude increase in U because C$^{+++}$
is going to higher stages of ionization.  All of these effects can
be seen in Figure 1 of Shields et al. (1995).  While it would
appear that C IV can indeed show the negative responsivity needed
for the Sparke (1993) model, this only occurs at a value of $U$
that is two orders of magnitude higher than the canonical value in
the BLR. Under such conditions the cloud is very optically-thin
and as we have noted in \S 3, the equivalent width of the lines
produced by the gas is too small.  Thus any gas with a negative
responsivity is going to produce a negligible contribution to the
total line flux.

We suggest that the differences in the slopes and widths of the
correlation functions are on the order of the errors in these
functions.  The noisier the measurements, the more peaked the
autocorrelation function is, and the line fluxes have higher
uncertainties than the continuum intensity. In the limit of
infinite noise one gets the sampling-window autocorrelation
function (see Gaskell \& Peterson 1987).  This effect is obvious
in the Mg\,II autocorrelation function in Figure 1 in Sparke
(1993) where there is a clear spike at the origin, but, in
general, the uncertainties in the correlation functions will mask
the narrow spike. Inspection of additional correlation functions
support this interpretation. In Figure 14 in Korista et al.
(1995), for example, the width of the optical-UV cross-correlation
function is narrower both at half maximum and at zero than the
autocorrelation function of the UV alone.  A negative responsivity
of the optical emission to changes in the UV is not a
physically-viable explanation of this. If the line autocorrelation
functions really are broader than the continuum autocorrelation
function (which we do not believe) one possible explanation could
be that the observed UV/optical continuum is itself reprocessed
from the photoionizing continuum. Multi-wavelength observations of
NGC\,7469 (Wanders et al. 1997, Collier et al. 1998) show that
there is such continuum reprocessing going on in at least
NGC\,7469. The failure of the NGC\,5548 H$\beta$ transfer function
to peak at zero (see Horne, Welsh, \& Peterson 1991), on the other
hand, is highly significant and must be real.  We suggest that the
explanation of this is one offered by Horne et al. (1991): that
the H$\beta$ emitting gas is not spherically symmetric.  Since the
Balmer-line profiles of NGC\,5548 show prominent structure and we
believe such structure is due to a disk component (Gaskell \&
Snedden 1999), we consider this very likely.
\subsection{An Optically-Thin Contribution to the General BLR?}
While we have argued that the VBLR is {\it not} predominantly
optically-thin, it would be highly unlikely if there were not at
least some optically-thin clouds in both the VBLR and the rest of
the BLR.  A real BLR doubtless consists of clouds with a wide
range of sizes -- perhaps with a fractal structure as suggested by
Bottorff \& Ferland (2001).  The change in line ratios as the
continuum varies probably provides the best evidence for this.
Wamsteker \& Colina (1986) found that the normal rapid rise of
C\,IV flux with continuum level in Fairall\,9 appeared to stop
when the continuum reached a high level.  The effect was less
pronounced for Ly$\alpha$.  They suggested that the effect was due
to clouds making a transition from being radiation-bounded to
being matter-bounded (i.e., optically-thin). This effect has been
thoroughly studied by Shields, Ferland, \& Peterson (1995).
Allowing for the lags in the C\,IV and Ly$\alpha$ emission removes
the abruptness of the effect, but the C\,IV/Ly$\alpha$ ratio still
declines at high luminosities.  Like Shields et al. (1995) we are
unable to reproduce this behavior with a purely optically-thick
model and we concur that an optically-thin contribution is needed
as well.
\section{CONCLUSIONS}
We conclude that rather than being optically-thin, the
high-velocity BLR gas is predominantly optically-thick.  We
believe this because optically-thin models fail to explain the
following properties of the high-velocity gas:
\begin{enumerate}
\item the equivalent widths, \item the Ly$\alpha$/H$\alpha$ ratio,
\item the Balmer decrement, \item the similarities of line
profiles, \item the strong variability of the high-ionization
lines, \item the C\,{\sc III}]/C\,{\sc IV} ratios, and \item the
strength of the $\lambda$1400 blend.
\end{enumerate}
If the high-velocity gas is predominantly optically-thick it is
thus not fundamentally different, in terms of optical depth, from
the lower-velocity gas.
\acknowledgments We are grateful to Giovanna Stirpe for making her
optical spectra available in machine-readable format.  We also
wish to acknowledge very helpful and detailed comments by Mark
Bottorff, Liz Klimek, Kirk Korista, Joe Shields, and Bev Wills.
This work has been supported in part by grant AR-05796.01-94A from
the Space Telescope Science Institute, which is operated by AURA,
Inc., under NASA contract NAS5-26555.

\clearpage
\begin{deluxetable}{lr}
\tablewidth{0pt} \tabletypesize{\small} \tablecaption{Galactic
Reddening} \tablehead{ \colhead{Object} & \colhead{E(B-V)}}
\startdata NGC 3783 & 0.120 \\ NGC 5548  & 0.011\\ PG 1351+64  &
0.014\\ Fairall 9   & 0.014 \\ PG 1211+143  & 0.034 \\ PG 1116+215
& 0.012 \\ B2 2201+31  &  0.015 \\ PKS 2251+11 &  0.064 \\
\enddata
\end{deluxetable}
\clearpage
\begin{deluxetable}{lrrrrrrr}
\tablewidth{0pt} \tabletypesize{\small}
\tablecaption{Ly$\alpha$/H$\alpha$ Line Ratios at Selected
Velocities corrected for Galactic Reddening} \tablehead{
\colhead{Object} & \colhead{-10000} & \colhead{-5000} &
\colhead{-2500}  & \colhead{0} & \colhead{2500}
& \colhead{5000} & \colhead{10000}\\
\colhead{} & \colhead{(km s$^{-1}$)} & \colhead{(km s$^{-1}$)} &
\colhead{(km s$^{-1}$)} & \colhead{(km s$^{-1}$)} & \colhead{(km
s$^{-1}$)} & \colhead{(km s$^{-1}$)} & \colhead{(km s$^{-1}$)}}
\startdata NGC 3783 &4.6 & 4.4    & 1.4 & 1.3 & 2.4 & 5.3 &
\nodata \\
NGC 5548  & \nodata  &   1.8    & 0.44 & 1.5    & 0.89 & 2.1 &
\nodata  \\
PG 1351+64  &   11.3    &   10.4  & 9.6 &   6.0  & 5.2 & 3.8
& 7.1 \\
Fairall 9   &   4.5 &   4.4    & 4.2 & 6.0 & 3.0 & 8.3 &
11.9 \\
PG 1211+143  &   2.0    &   4.8   & 3.1 & 2.6 & 2.6 & 5.6 &
5.4 \\
PG 1116+215 &   16.1  &   16.4 & 8.7 & 3.4 & 6.1 & 15.0 &
\nodata \\
B2 2201+31  &   7.0   &   5.6 & 7.3 & 3.0 & 4.0 & 4.7 & 8.2
\\
PKS 2251+11 &   4.7    &   3.2  & 2.1 & 1.8 & 1.4 & 2.3 &
5.4\\
\enddata
\end{deluxetable}
\clearpage
\begin{deluxetable}{lrrrrrrr}
\tablewidth{0pt} \tabletypesize{\small}
\tablecaption{H$\alpha$/H$\beta$ Line Ratios at Selected
Velocities corrected for Galactic Reddening} \tablehead{
\colhead{Object} & \colhead{-10000} & \colhead{-5000} &
\colhead{-2500}  & \colhead{0} & \colhead{2500}
& \colhead{5000} & \colhead{10000}\\
\colhead{} & \colhead{(km s$^{-1}$)} & \colhead{(km s$^{-1}$)} &
\colhead{(km s$^{-1}$)} & \colhead{(km s$^{-1}$)} & \colhead{(km
s$^{-1}$)} & \colhead{(km s$^{-1}$)} & \colhead{(km s$^{-1}$)}}
\startdata NGC 3783 & 0.89   & 1.2 & 2.7   & 3.6 & 2.8   & 1.6 &
\nodata \\ NGC 5548 & 1.3 & 2.2 & 3.7 & 4.0 & 4.4 &
2.9 & \nodata \\
PG 1351+64 & \nodata   & 2.0 & 2.8 & 3.3 & 2.8 & 2.6 & 2.1
\\
Fairall 9 &  2.0   &   3.8 & 3.1 & 5.0 & 4.5 & 2.6 & 1.1\\
PG 1211+143 & 1.1  & 3.8 & 2.2 & 3.3 & 2.4 & 2.0 & \nodata
\\
PG 1116+215 & 0.69 & 2.3    & 2.2 & 3.1 & 2.6 & 1.9 & \nodata
\\
B2 2201+31 & 1.8 &   4.0   & 3.3 & 3.8 & 4.0 & 3.4 & 2.0\\
PKS 2251+11 & \nodata  & 3.8  & 4.4 & 4.5 & 4.2 & 2.4 &
1.6\\
\enddata
\end{deluxetable}
\clearpage
\figcaption[LaHa2.eps, Laha1.eps]{The L$\alpha$ profile in PG
1116+ 215, with no de-blending, compared with the H$\alpha$
profile in the same object.  [N II] has been removed from the
H$\alpha$ profile. Figure 1\emph{(a)} shows the two lines
normalized to the same peak; Figure 1\emph{(b)} shows them
normalized so that the short-wavelength wings match.  The profile
differences at +5000 to +10,000 km s$^{-1}$ mostly likely have a
significant contribution from N V and Si II emission.}
 \figcaption[avelaha.ps, avehahb.ps]{Observed hydrogen
line ratios. {\it (a)} Average values of Ly$\alpha$/H$\alpha$
across the line profile for our sample are shown as boxes. These
points are corrected for Galactic reddening only. The dashed curve
shows Ly$\alpha$/H$\alpha$ after an additional intrinsic
dereddening corresponding to E(B-V) $\approx 0.1$. {\it (b)} Same
as {\it (a)} but for the H$\alpha$/H$\beta$ ratio. \label{fig1}}
\figcaption[thinhydro.ps,hydro1z.ps, logtau_0_hydro.ps] {Grids of
modeled hydrogen line ratios. {\it (a)} An optically-thin
photoionization model. Solid lines show contours of
Ly$\alpha$/H$\alpha$; dashed lines show H$\beta$/H$\alpha$. We use
H$\beta$/H$\alpha$ here to avoid confusion with the values for
Ly$\alpha$/H$\alpha$. {\it (b)} Same as {\it (a)}, but for
optically-thick gas. The average ratios of our sample at $\pm
5000$ km s$^{-1}$ and 0 km s$^{-1}$ fall inside the marked
ellipse. {\it (c)} Same as {\it (a)}, but for a model that stops
at $\tau_{912} = 1$, yielding the minimum reasonable $W_{\lambda}$
for the Ly$\alpha$ broad emission line component. \label{fig2}}
\figcaption[thincarbon.ps,carbon1z.ps, logtau_0_carbon.ps] {Grids
of modeled carbon line ratios. {\it (a)} An optically-thin
photoionization model. Solid lines show contours of
Ly$\alpha$/C\,{\sc IV}; dashed lines show C\,{\sc III}]/C\,{\sc
IV}. {\it (b)} Same as {\it (a)}, but for optically-thick gas.
{\it (c)} Same as {\it (a)}, but for a model that stops at
$\tau_{912} = 1$, yielding the minimum reasonable $W_{\lambda}$
for the Ly$\alpha$ broad emission line component. \label{fig3}}
\newpage
\plotone{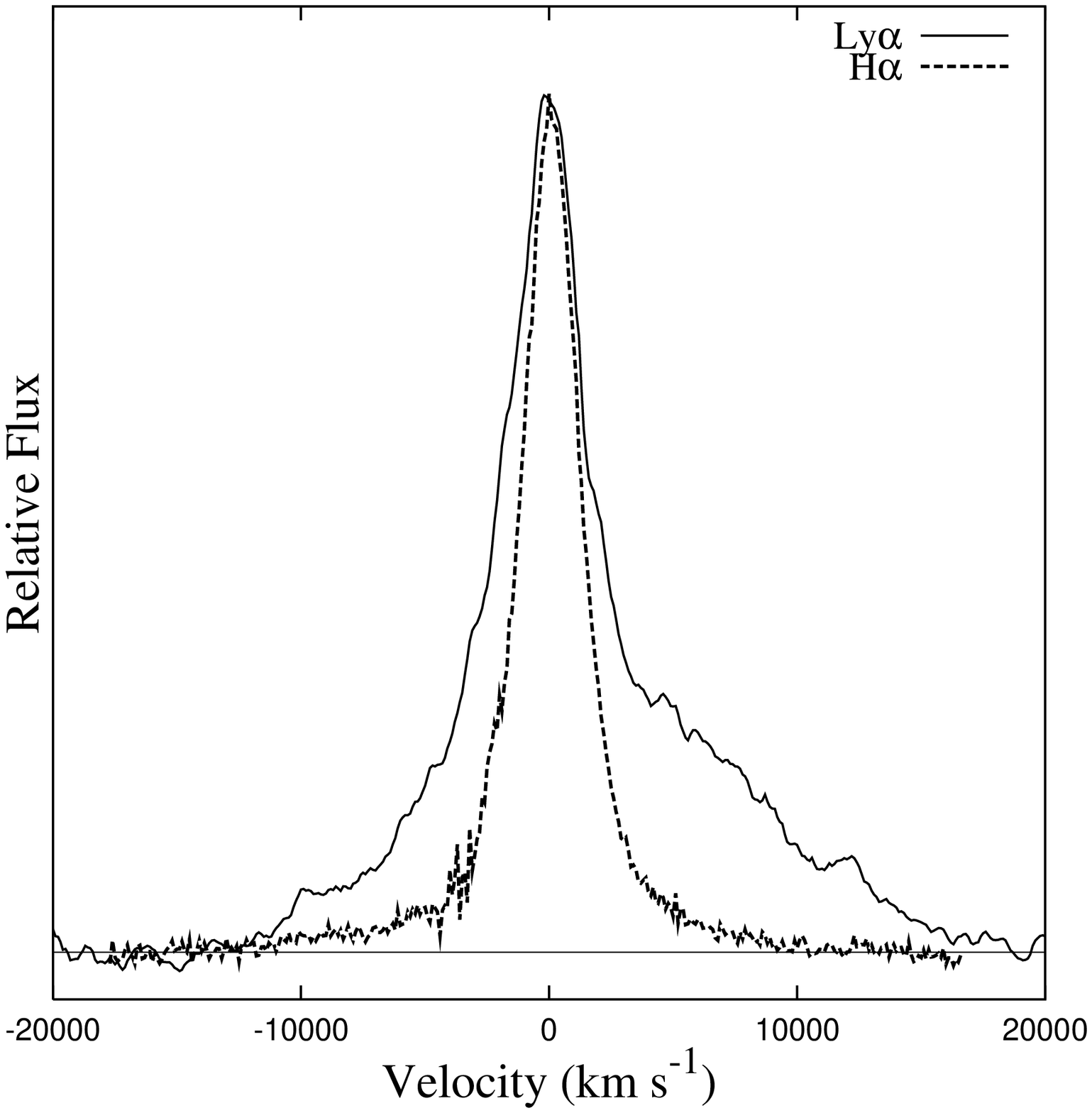}
\newpage
\plotone{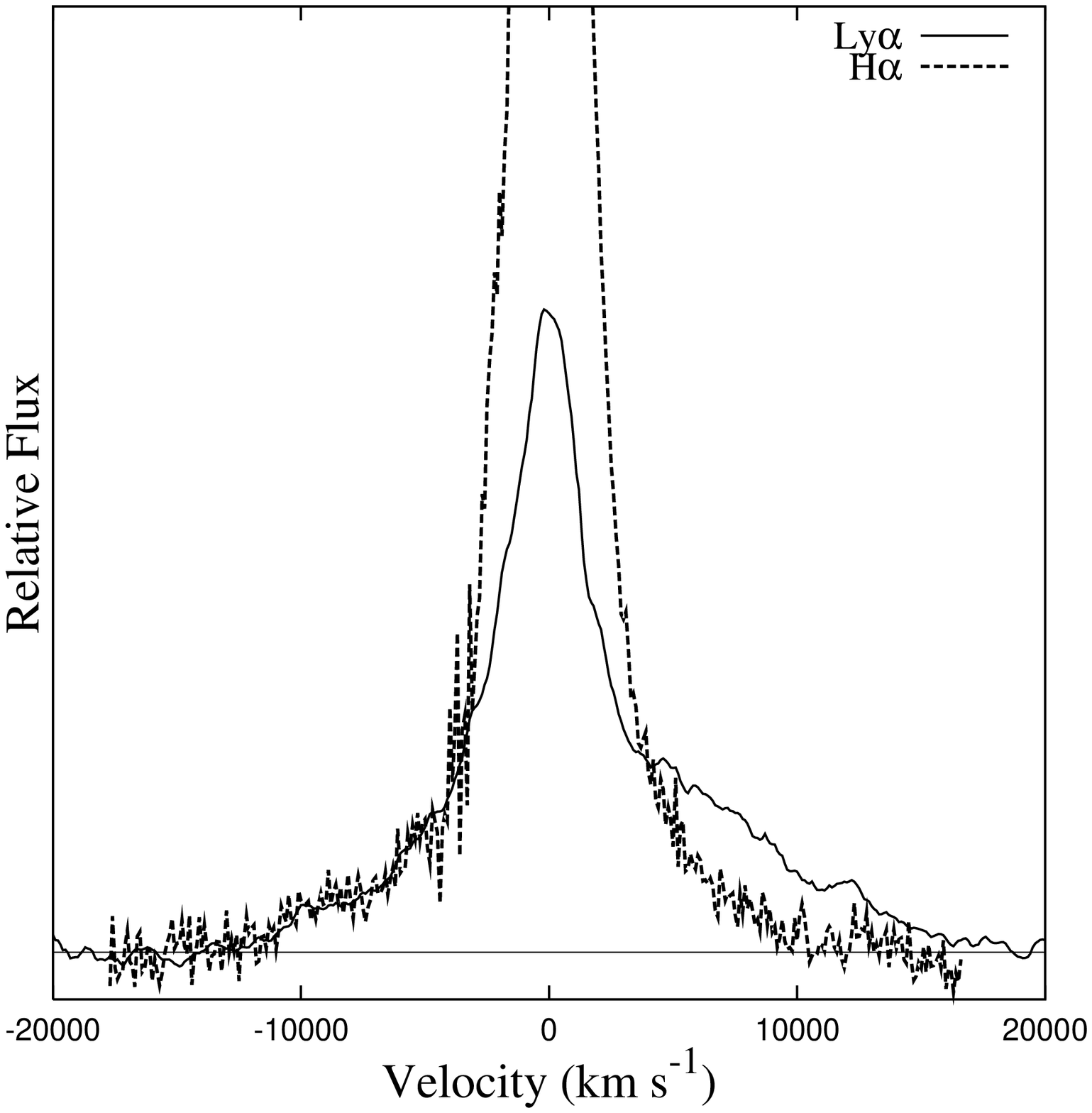}
\newpage
\plotone{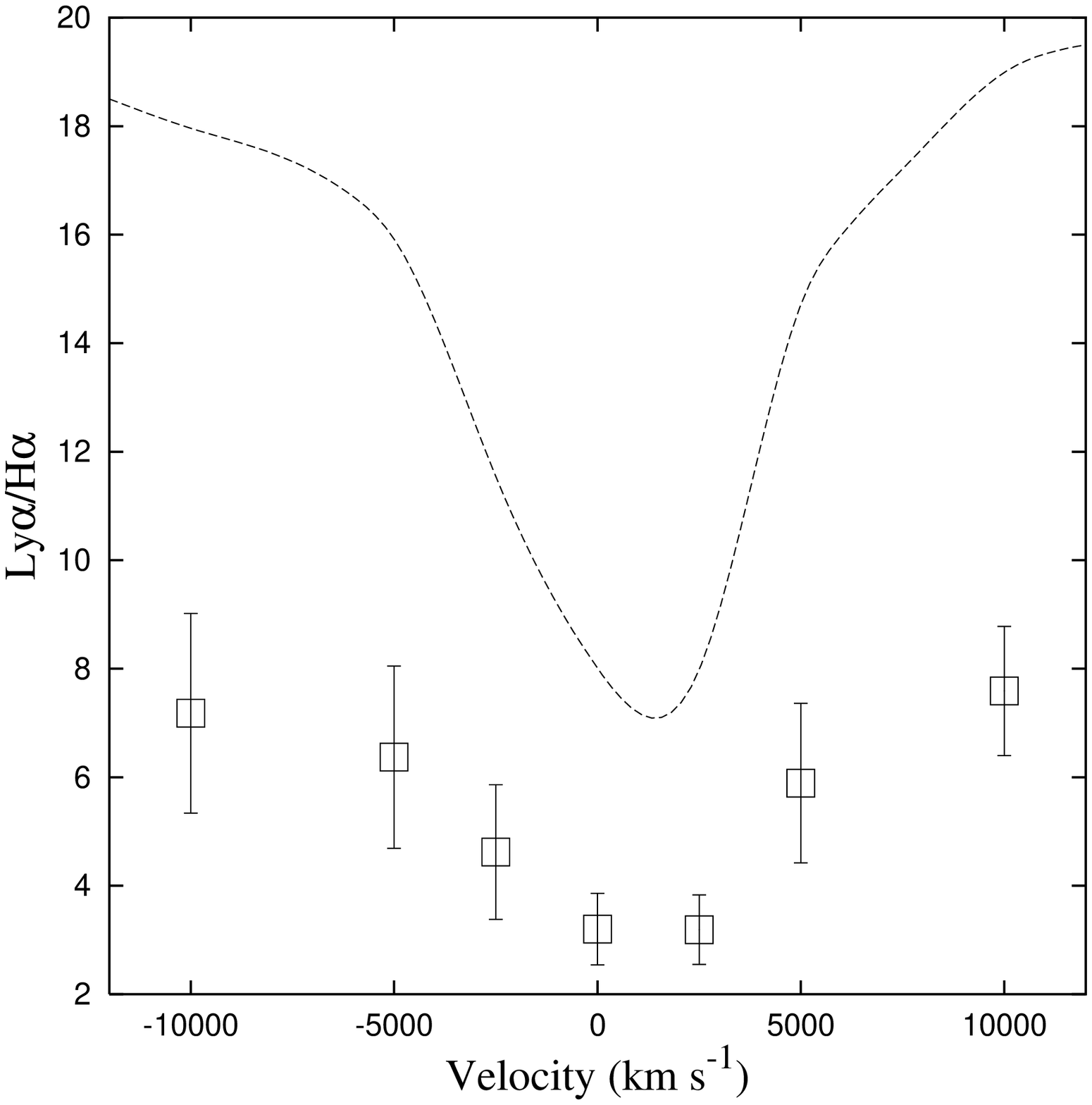}
\newpage
\plotone{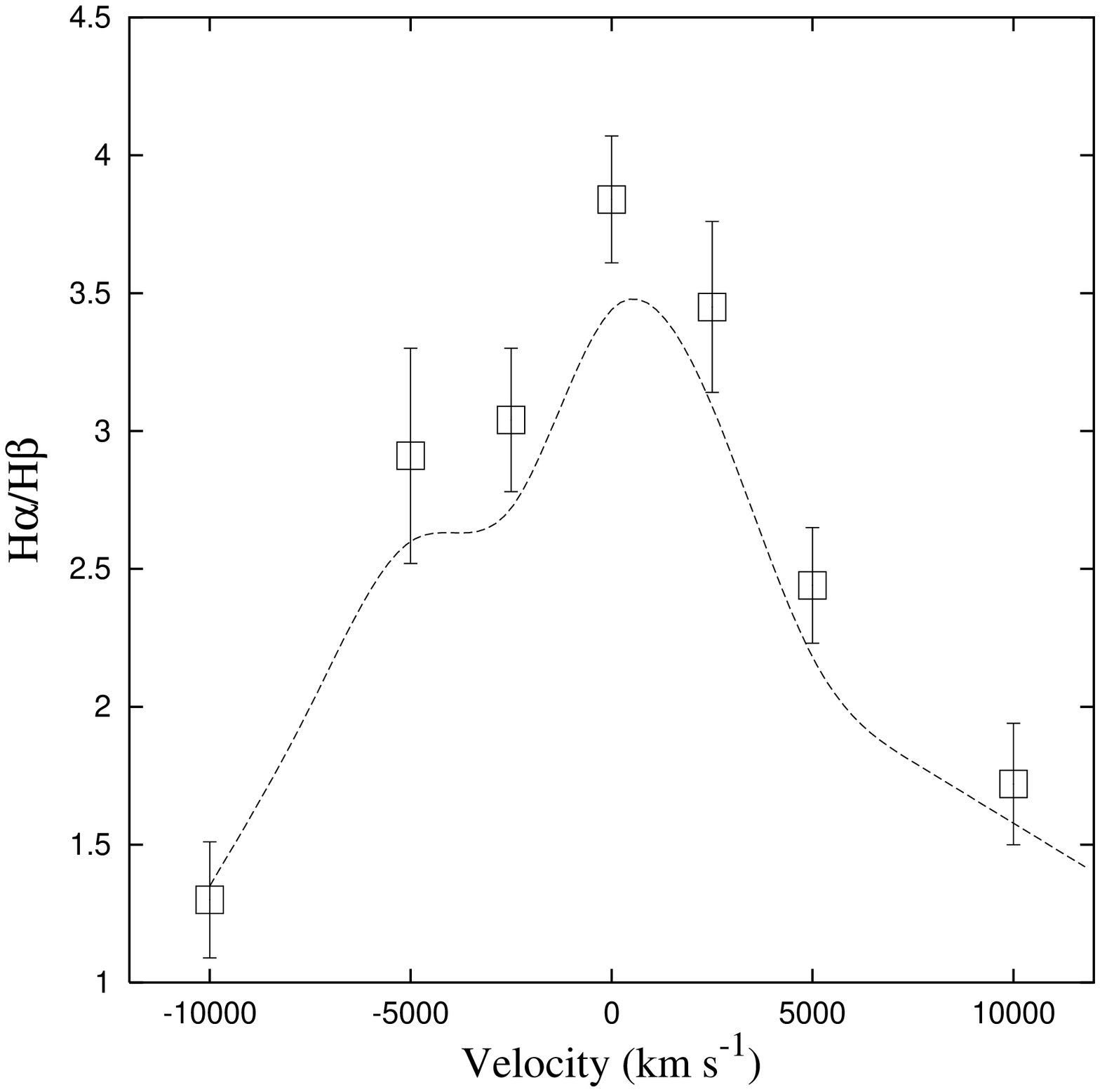}
\newpage
\plotone{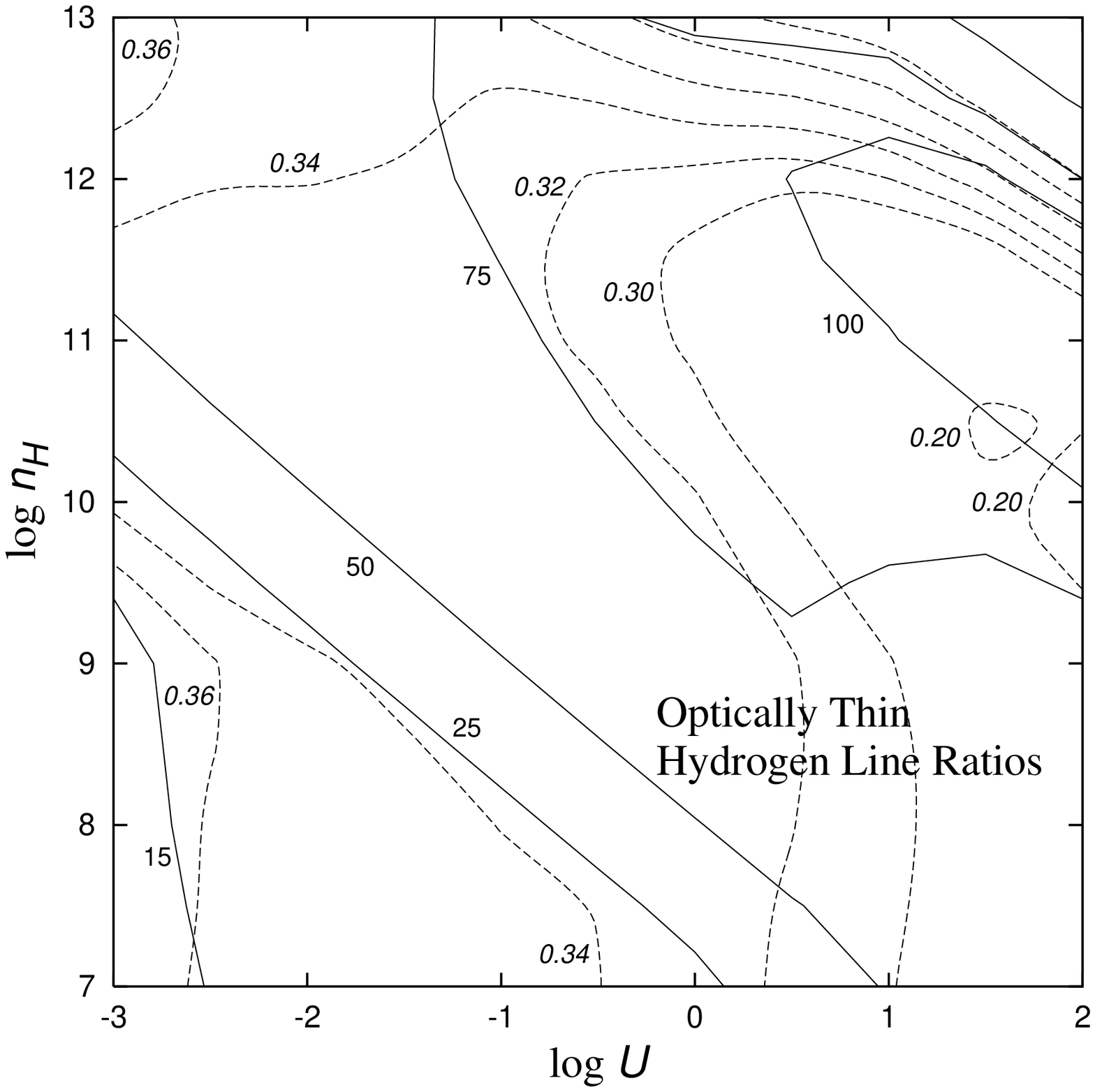}
\newpage
\plotone{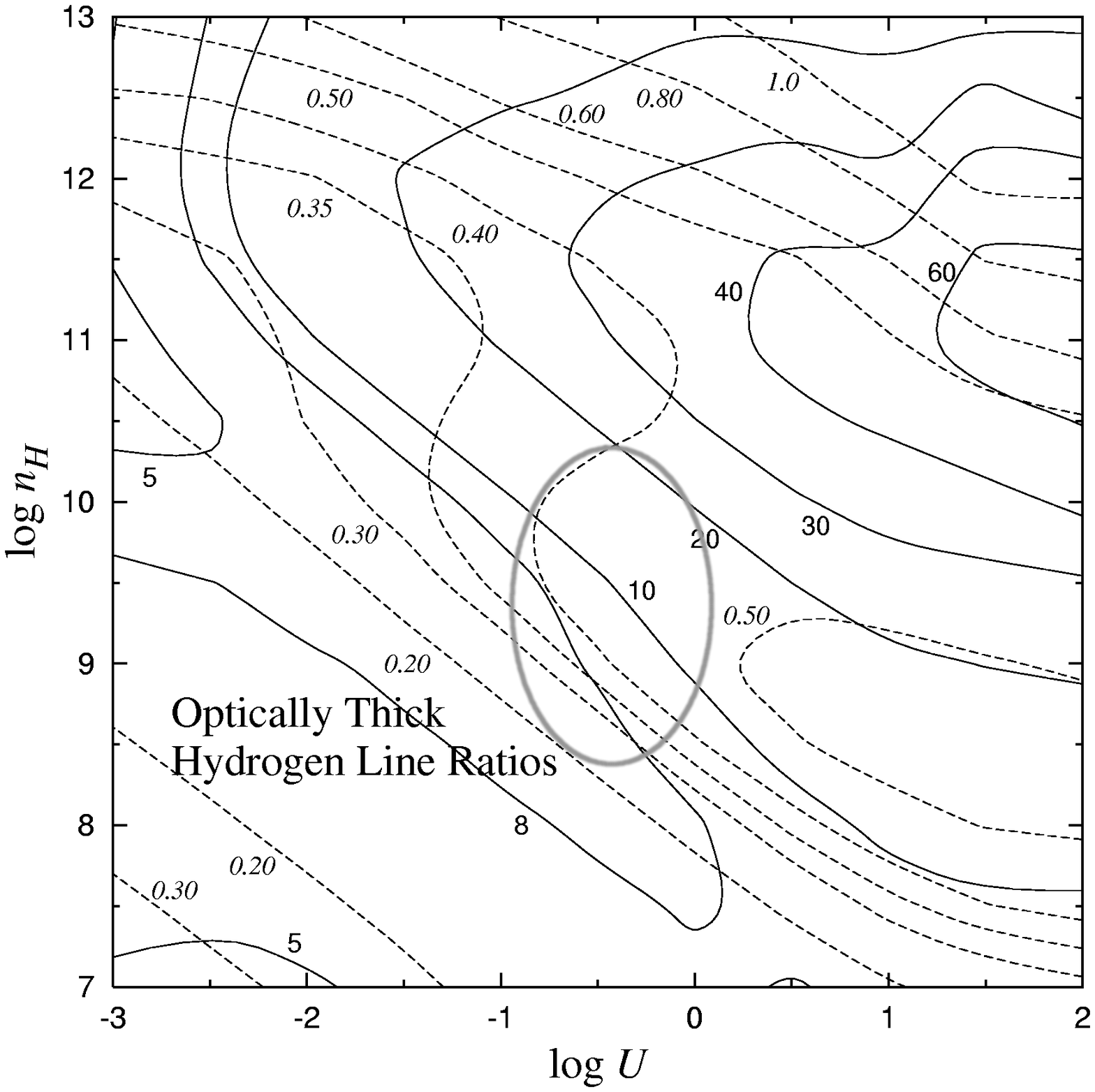}
\newpage
\plotone{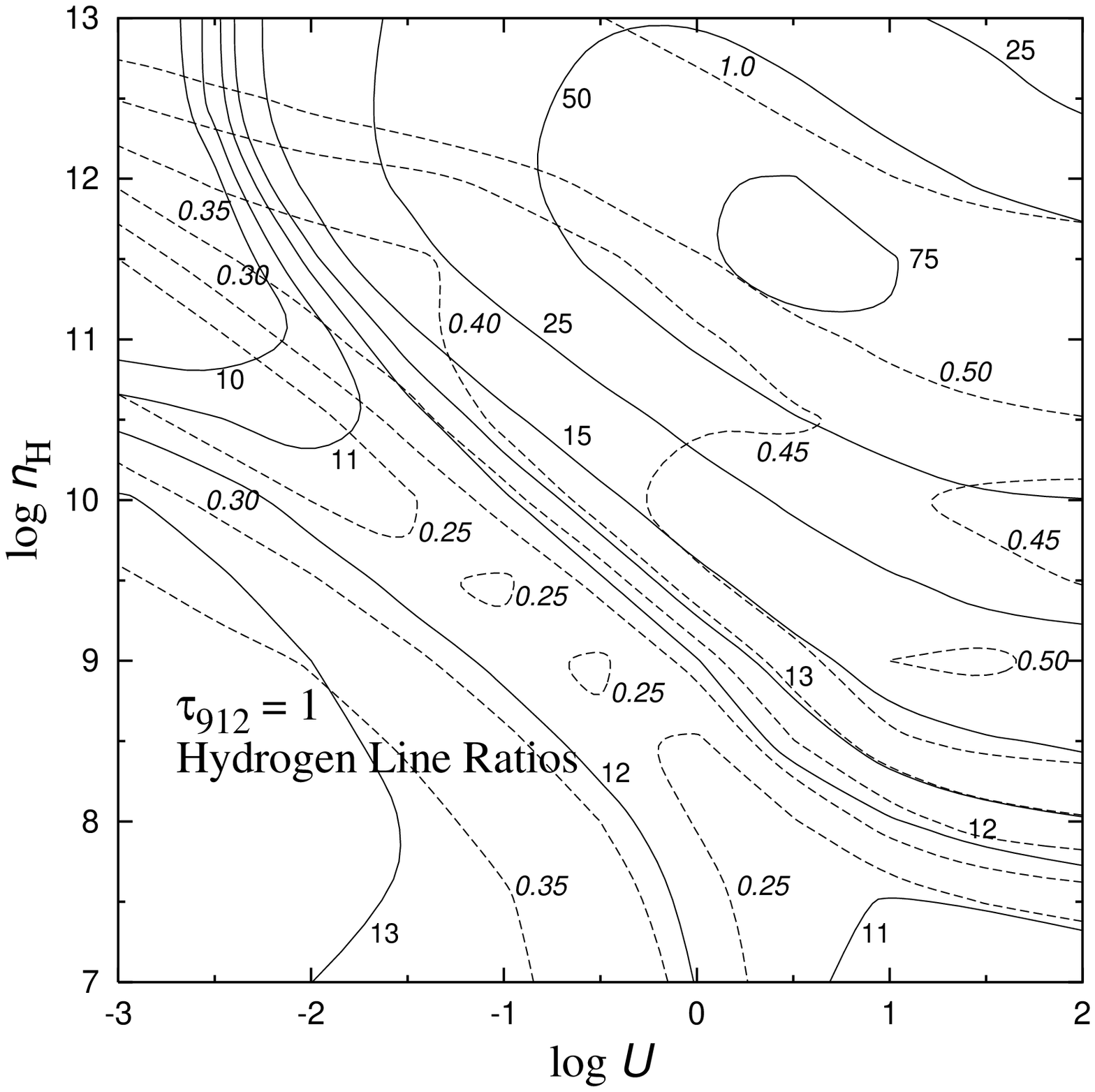}
\newpage
\plotone{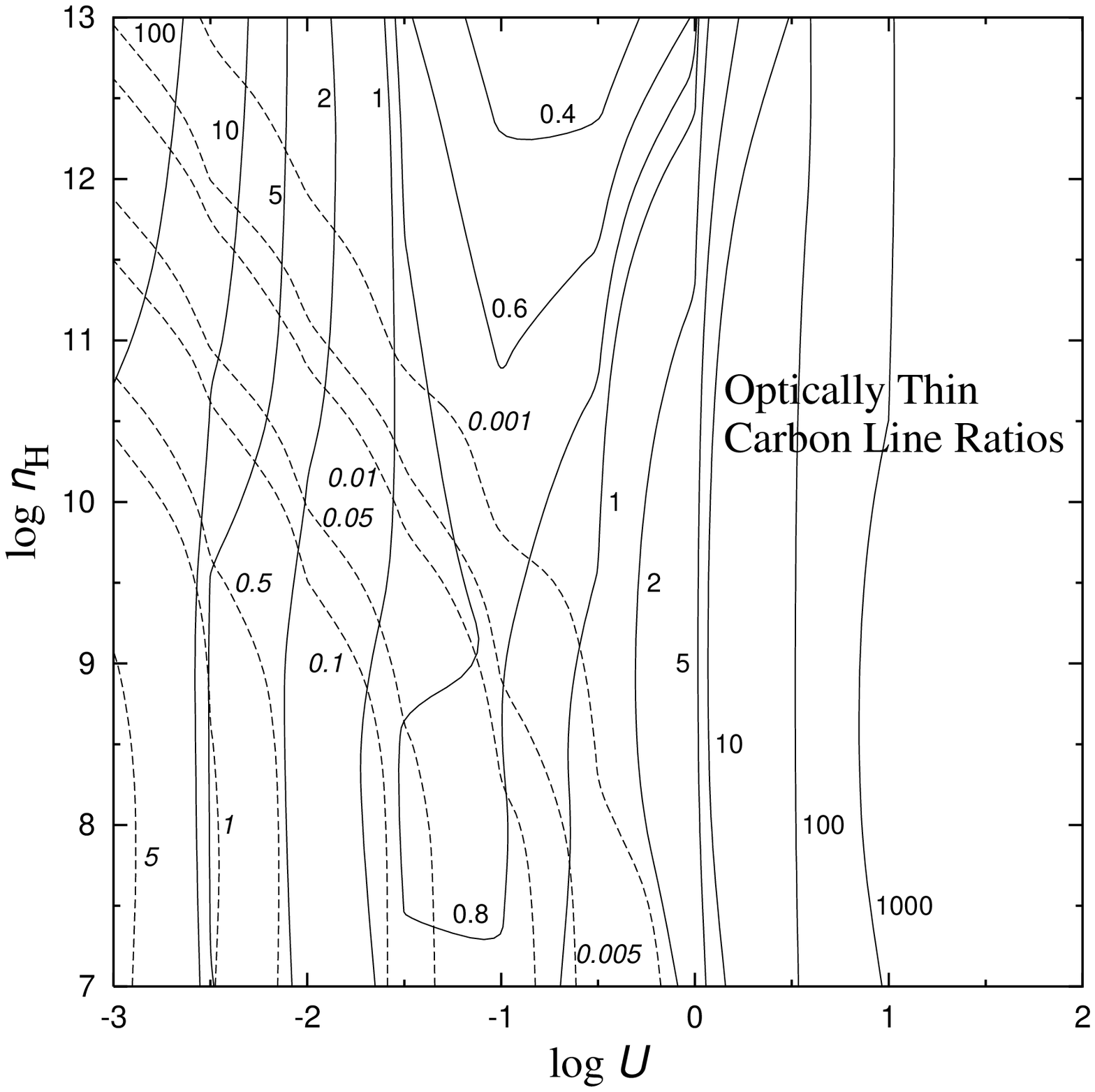}
\newpage
\plotone{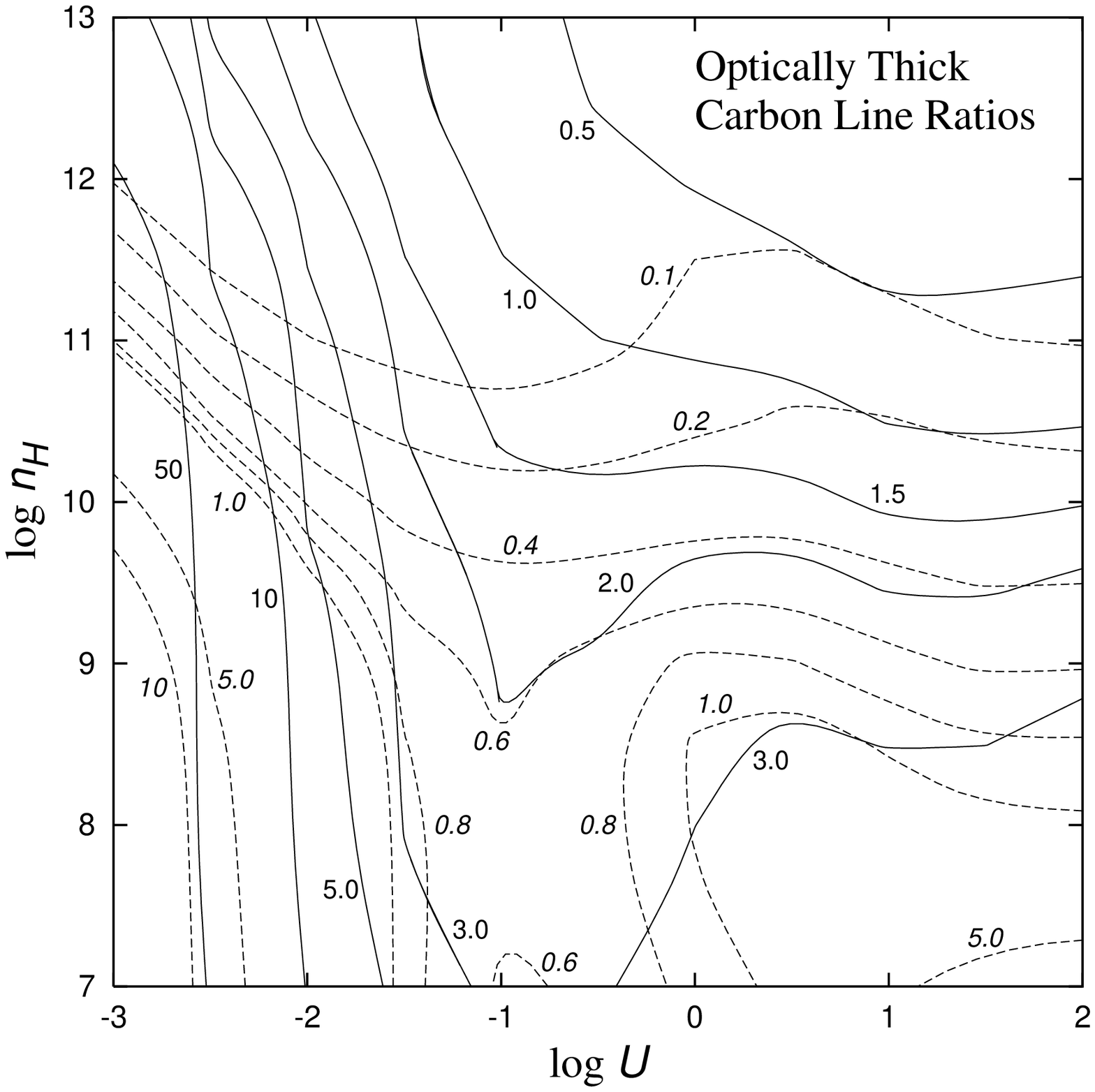}
\newpage
\plotone{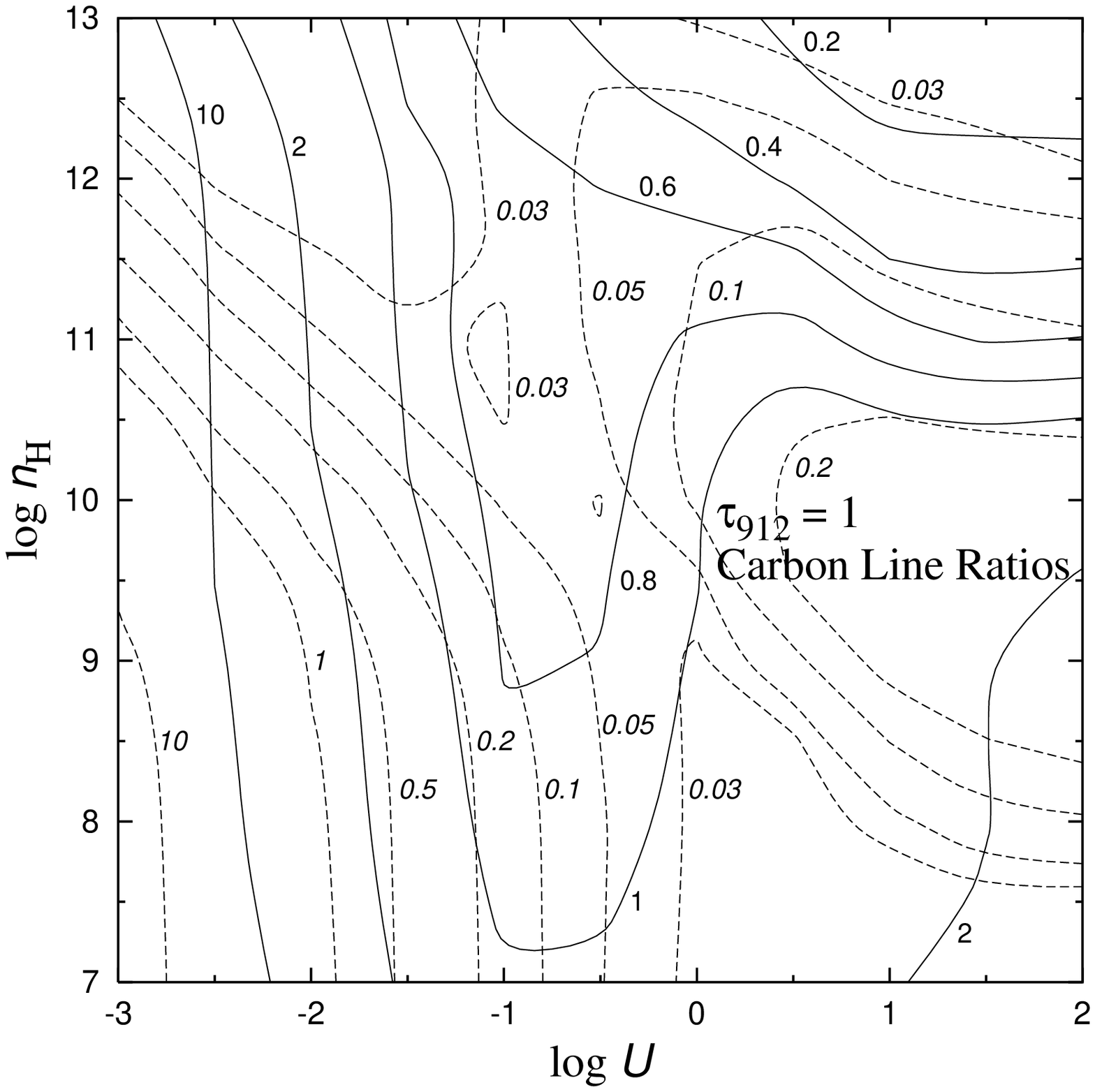}
\newpage
\end{document}